# THE LARGE-SCALE DISTRIBUTION OF QUASARS

B. V. Komberg[1], A. V. Kravtsov[2], and V. N. Lukash[1]

[1]*Astro Space Center, Lebedev Physical Institute, Moscow*
[2]*Physics Department, Moscow State University, Moscow*

**Abstract.** We report the investigation of spatial distribution of quasars using two different methods: the statistical study by means of 2-point correlation function and direct search for structures such as previously reported Large Quasar Groups.

Having analyzed the combined sample of eight homogeneous quasar surveys (totalling near 2200 objects), we detected strong clustering at scales $r \leq 20h^{-1}$ Mpc[1] which seems, however, to be redshift dependent. Using reasonable selection criteria we have sought for Large Quasar Groups (LQG) similar to those which were earlier found observationally. We discuss the properties of the detected groups and their implications for cosmology. We also discuss the suitability of the present data for getting the reliable statistical inference on quasar clustering.

## 1 Introduction.

The rapid progress in the study of large-scale structure in nearby regions using galaxies and their clusters offered the possibility to compare the predictions of various structure formation theories directly with observations and to reject those models which do not fit them well. In this framework, the study of spatial distribution of quasars is the unique possibility of probing these theories at high redshifts, i.e., when the Universe was much younger, and defining, thus, not only the properties of structure but also its "date of birth" and evolution.

Here we report the results of quasar distribution investigations using both the statistical method and the direct search for structures in the present data. In Section 2 we will focus on the correlation analysis of a combined sample which consists of eight homogeneous quasar surveys. In Section 3 we present results of the search for quasar groups. In Section 4 we discuss the results and draw our conclusions.

## 2 The investigation of quasar clustering using two-point spatial correlation function.

Eight published homogeneous surveys of quasars totalling about 2200 objects are used in our work (for references and detailed description of data see [2]). That is so far the largest sample ever investigated. We use the conventional method of two-point spatial correlation function. In order to reproduce the unique selection function for every survey (manifested in its redshift distribution) we used *scrambling* technique (see [1] and references therein) to create the random comparison catalogs. This technique is the permuting of redshifts between objects randomly while their celestial coordinates remain the same. Then the correlation function and its errors

---
[1]Unless otherwise stated, we assume $H_0 = 100h$ and $q_0 = 0.5$

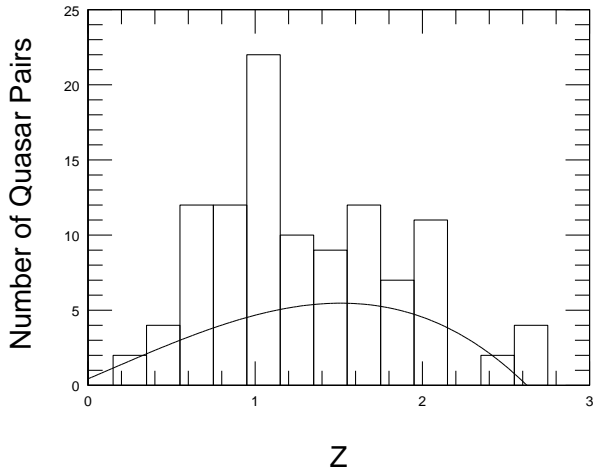

Figure 1: The close pair statistics versus redshift.

were calculated through the well-known estimator:

$$\xi(r) = \frac{N_{obs}}{N_{poiss}} - 1, \quad \Delta\xi = \sqrt{\frac{1+\xi}{N_{poiss}}},$$

where $N_{obs}(r)$ is the observed number of pairs with the distance between $r$ and $r + dr$ in the survey under consideration, $N_{poiss}(r)$ is the corresponding average number of pairs in the random samples which were created one hundred times and $r$ is comoving separation between quasars in megaparsecs. Having applied this method to our data we obtained the following results: (i) a strong clustering was detected at scales less than $20h^{-1} Mpc$. (ii) the correlation function for the total sample can be expressed at these scales as a power law $\xi(r) = (r/r_0)^\gamma$ with $r_0 = 6.0 \pm 1.2h^{-1}$ Mpc when $\gamma = -1.8$. The correlation amplitude, however, was found to be redshift dependent (see Fig.1). We argue [2] that pair statistics depends strongly on the limit magnitude of survey (the deeper survey is, the more its quasar density is, related directly to statistics of pairs at small separations). In fact, we can estimate the statistics before the analysis, calculating the number of pairs with a given separation in case the quasars were distributed randomly (see [2] for quantitative analysis). The distribution of close pairs versus redshift is drawn in Fig.1. Here the smooth curve represents the polinomial approximation of the distribution of poissonian pairs. Note that the ratio $N_{obs}$ to $N_{poiss}$ increases as redshift decreases in the interval $z \sim 1 - 2$, though, the statistics is still poor to draw more robust conclusions.

## 3    The search and investigation of Large Quasar Groups.

Though, no attempts were made to search systematically for structures in quasar distribution before[1], we stress [4] that such study may potentially be of great im-portance for our understanding of structure evolution and formation. Moreover, now we have observational evidence for existence of such structures. In [5, 6] the discovery of two groups of QSO was reported. The groups were found as an unusual enhancements of quasar number density (first group contains 23 QSOs, while second — 13) and were claimed to have sizes $\sim 100h^{-1}$ Mpc and clumpy inner structure. So a first step could be to check the existing quasar data for similar groups.

---

[1] The only attempt [3], however, not a systematic one, was unsuccsessful, probably because of the insufficient data

Here we briefly describe our search strategy and method of detecting of high quasar density sites which we call Large Quasar Groups (LQG)[2]. First we studied the Catalog of quasars [7] which contains more than 6000 objects. Here it is necessary to say that we are conscious of its shortcomings such as, for example, incompleteness, but the point is that we are not using the catalog in any statistical study but considering it just as a complete survey of literature. We apply the well-known clusterisation algorithm with a number of clusterisation radii to obtain the lists of formal clusters with sizes and quasar number densities computed for each cluster. We select then "candidate" groups using following selection criteria: (i) the cluster must contain at least ten objects; (ii) the number density of quasars in clusters must be at least twice the density of field QSOs at the same $z$; (iii) the objects in cluster must have reliable redshifts. It was found that all groups from the final list come from deep homogeneous surveys. This fact makes it possible to estimate an empirical probability for the group to be random using the survey from which it originates. We have created one thousand random catalogs on the basis of the original survey and applied the same procedure to select groups. Then the probability was estimated just by counting events of detection of similar groups in these random samples. The properties of the found groups are summarized in the Table 1. Here, in the first column, we give the identification number for the group[1], in the second — the number of quasars, in the third, in the fourth and in the fifth — the average redshift, size along the redshift and number density correspondingly, and in the last column — the estimated probability to be random[2].

## 4  Discussion and conclusions.

We have obtained, therefore, the following results. First, quasars seem to have the correlation properties similar to those of galaxies and galaxy clusters (in sense of the same power law form of correlation function). There are also the indications that the amplitude of correlations evolves with cosmic time. Second, we have found evidences for existence of structures in quasar distribution of typical size $\sim 100h^{-1}$ Mpc. The number of such structures or quasar groups found in our analysis is considerably larger than the number of groups known before (we have identified 11 new groups while only two were known). It seems then, that quasar groups are more common fenomenon than it was previously thought [6]. We argue [4] that the majority of QSOs at $z \leq 2-3$ form in merging and interacting galaxies produced during the first generation of matter crossings which occur in galaxy protoclusters. Groups of quasars may, therefore, belong to concentrations of young clusters of galaxies, and thus indicate the locations of high-density regions which develop later into quasi-linear systems like the local Great Attractor and superclusters. The study of these early structures provide a natural way to trace the matter distribution at high redshifts and may offer a possibility to determine the spectrum of primordial perturbations at scales encompassing both clusters and superclusters.

---

[2] the detailed description of the method will be published elsewhere
[1] LQG 9 was found before [5], and CC-group is the group found in [6] added for comparison
[2] for the first two LQG we couldn't estimate the probability because they contain QSOs from different surveys

Table 1: Summary of LQG analysis.

| Group | Number of QSOs | Redshift | $R_z$, $h^{-1}Mpc$ | Density, $\times 10^{-5} h^3 Mpc^{-3}$ | Probability to be random |
|---|---|---|---|---|---|
| LQG 1 | 12 | $\sim 0.6$ | 96 | 3.3 | – |
| LQG 2 | 12 | $\sim 0.8$ | 111 | 3.3 | – |
| LQG 3 | 14 | $\sim 1.3$ | 123 | 3.0 | 0.04 |
| LQG 4 | 14 | $\sim 1.9$ | 104 | 4.9 | $< 10^{-3}$ |
| LQG 5 | 13 | $\sim 1.7$ | 146 | 4.0 | 0.01 |
| LQG 6 | 10 | $\sim 1.5$ | 94 | 7.6 | 0.009 |
| LQG 7 | 10 | $\sim 1.9$ | 92 | 4.6 | 0.2 |
| LQG 8 | 12 | $\sim 2.1$ | 104 | 5.73 | |
| LQG 9 | 25 | $\sim 1.1$ | 66 | 5.1 | $< 10^{-3}$ |
| LQG 10 | 17 | $\sim 1.9$ | 164 | 2.3 | 0.01 |
| LQG 11 | 10 | $\sim 0.7$ | 157 | 2.4 | 0.08 |
| LQG 12 | 17 | $\sim 1.2$ | 155 | 3.0 | 0.05 |
| CC-group | 13 | $\sim 1.3$ | 154 | 0.45 | 0.03 |

**Aknowledgements.** It is pleasure to thank the staff of theoretical department of Astro Space Center for discussions, M. Stepantsov and B. Blinov for the help in preparing of the manuscript. Our work was partly supported by the Russian Foundation for Fundamental Research (the project code 93-02-2929) and by the ESO C&EE grants (NN A-01-152, A-03-003)